\documentclass[english,a4paper,manuscript]{revtex4}
\usepackage[T1]{fontenc}
\usepackage[latin1]{inputenc}
\usepackage{amstext}

\makeatletter
\@ifundefined{textcolor}{}
{%
 \definecolor{BLACK}{gray}{0}
 \definecolor{WHITE}{gray}{1}
 \definecolor{RED}{rgb}{1,0,0}
 \definecolor{GREEN}{rgb}{0,1,0}
 \definecolor{BLUE}{rgb}{0,0,1}
 \definecolor{CYAN}{cmyk}{1,0,0,0}
 \definecolor{MAGENTA}{cmyk}{0,1,0,0}
 \definecolor{YELLOW}{cmyk}{0,0,1,0}
 }

\makeatletter

\makeatletter

\makeatletter

\makeatletter

\makeatletter

\makeatletter

\usepackage{geometry}

\makeatother

\makeatother

\makeatother

\makeatother

\makeatother

\makeatother

\makeatother

\usepackage{babel}

\begin{document}

\title{Remarks on relativistic kinetic theory to first order in the gradients}

\author{A. L. Garcia-Perciante$^{2}$, A. Sandoval-Villalbazo$^{1}$}

\address{$^{1}$Depto. de Fisica y Matematicas, Universidad Iberoamericana,
Prolongacion Paseo de la Reforma 880, Mexico D. F. 01219, Mexico}

\address{$^{2}$Depto. de Matematicas Aplicadas y Sistemas, Universidad Autonoma
Metropolitana-Cuajimalpa, Artificios 40 Mexico D.F 01120, Mexico.\\
 }
\begin{abstract}
In this paper we emphasize some conceptual points related to the kinetic
foundations of relativistic hydrodynamics. We summarize previous work
and focus on the construction of the heat flux from a kinetic theory
point of view. A thorough discussion addressing aspects concerning
stability, causality and the construction of an appropriate stress-energy
tensor is included.
\end{abstract}
\maketitle

\section{Introduction}

The first solid theory formulated to deal with irreversible processes
is due to Lars Onsager \citep{Onsager1} \citep{Onsager2} and its
first version was published almost eighty years ago. Extensions and
modifications of his method were made by Casimir \citep{Casimir1}
and almost simultaneously by Meixner \citep{Meixner1}, all of which
gave rise to what we presently know as linear irreversible thermodynamics
(LIT). However, in 1940, before LIT was finally accomplished, C. Eckart
published three papers on irreversible thermodynamics \citep{Eckart1},
the first two more or less along the lines of LIT and the third one
dealing with the relativistic irreversible thermodynamics of a simple
fluid. Thus arises the question: is this work the correct extension
of LIT to relativistic systems? The answer has been subject to continuous
debate for over forty years and this is precisely the motivation of
this work. It is important to point put that no new results are included
in this publication. Following a very much appreciated suggestion
from an anonymus referee, this review article has been written as
a thorough discussion of some key theoretical aspects that lie beneath
recent results that can be found in the literature. Some basic equations
are included for the sake of clarity.

It has recently been shown \citep{Nos1} that the source of the generic
instabilities of a relativistic fluid \citep{HL} is the misuse of
the constitutive equation for the heat flow first proposed by Eckart
in 1940 \citep{Eckart1}. In that work it is clearly shown how the
acceleration term in the expression

\begin{equation}
J_{[Q]}^{\nu}=-\kappa h_{\mu}^{\nu}\left(T^{,\mu}+\frac{T}{c^{2}}\dot{u}^{\mu}\right)\label{eq:fr}\end{equation}
is the one that leads to the exponential growth of fluctuations around
the equilibrium state in the linearized system of relativistic transport
equations. Here $J_{[Q]}^{\nu}$ is the heat flux, $\kappa$ a relativistic
thermal conductivity, $h^{\mu\nu}$ is the usual spatial projector
defined as $h^{\mu\nu}=g^{\mu\nu}+\frac{u^{\mu}u^{\nu}}{c^{2}}$ for
a $\left(+++-\right)$ signature, $T$ is the local temperature, $c$
the speed of light and $u^{\mu}$ the hydrodynamic four velocity.
The covariant derivative is denoted by a semicolon, the ordinary one
as a comma and a dot implies a total proper-time derivative such that
$\dot{a}^{\nu}=u^{\mu}a_{;\mu}^{\nu}$. In this work we point out
two major drawbacks of this constitutive equation and thus proceed
to address the matter from the kinetic theory point of view. We repeat
some key equations leading to the constitutive relation obtained from
the relativistic Boltzmann equation as well as the ones needed in
order to address both the violation of the linear regression assumption
and the causality of the first order (in the gradients) theory.

To accomplish this task in Section II we start by recalling the basic
elements of Onsager's theory in order to clearly expose the shortcomings
introduced in first order theories by Eq. (\ref{eq:fr}). In Section
III we outline the calculation of Ref. \citep{physica A} where a
constitutive equation for the heat flux is derived by solving the
Boltzmann kinetic equation for a relativistic ideal gas in a BGK-like
approximation. In Section IV we briefly summarize the key steps and
arguments favoring the kinetic theory result and showing its consistency
with linear irreversible thermodynamics. Finally, the key aspects
of the work are summarized and discussed in Section V.

\section{Generic Instabilities and Onsager's Hypothesis}

The constitutive equation (\ref{eq:fr}), arises from phenomenologically
enforcing the second law of thermodynamics to a system in which the
stress-energy tensor includes relativistic corrections terms proportional
to the heat flux. Such tensor is obtained as the most general decomposition
of a second rank tensor in this framework. This procedure is the standard
one in non-equilibrium thermodynamics, however it leads to a consitutive
equation which is at odds with its basic tenets in two senses. First,
according to this theory, fluxes must be coupled to the forces given
by the gradients of the intensive variables. This requirement is clearly
not met by the second term in Eq.(\ref{eq:fr}), the acceleration
is neither an independent variable nor a thermodynamic force. The
second problem, as mentioned above, is the fact that it leads to a
violation of the linear regression assumption, a behavior usually
referred to as {}``generic intability''. As was pointed out in Ref.
\citep{Nos1}, Onsager's hypothesis states that spontaneous fluctuations
of the state variables around the equilibrium state, whose origin
is purely microscopic, should relax following the linearized equations
for such variables. Thus, the exponential growth obtained by Hiscock
and Limdblom is really a violation of this assumption which is key
in the construction of irreversible thermodynamics and in particular
in the proof of Onsager's reciprocity relations. Indeed, the theory
of irreversible processes developed by Onsager seeks the establishment
of a connection between these processes and the spontaneous fluctuations
that due to their microscopic nature appear in the thermodynamic variables
of equilibrium systems. To do so he used the local equilibrium assumption
(LEA) \citep{DeGroot2} \citep{GC1} by assuming the thermodynamic
state of a system to be determined through a set of extensive variables
$\left(\alpha_{1},\alpha_{2},...,\alpha_{n}\right)=\vec{\alpha}$
(energy, volume, particle number, etc) with which all thermodynamic
variables, in particular the entropy $S$, may be defined. These $\alpha^{\prime}s$
are further redefined so that they vanish in the equilibrium state,
$S\left(\vec{\alpha}\right)_{Eq}=S\left(\vec{0}\right)=S_{0}$ and
thus, the thermodynamic forces acting on the system to seek equilibrium
are:\begin{equation}
X_{i}=\left(\frac{\partial S}{\partial\alpha_{i}}\right)_{j\neq i}\label{one}\end{equation}
while the fluxes are simply the time derivatives of the $\alpha^{\prime}s$.
The essential assumption in Onsager's formulation is that the fluxes
are linearly related to the forces $X_{i}$ so that \begin{equation}
\frac{d\alpha_{i}}{dt}=\sum_{j=1}^{n}L_{ij}X_{j}\label{two}\end{equation}
Further, to determine the force $X_{j}$ it is assumed that $S\left(\vec{\alpha}\right)$
can be expanded in a Taylor series around equilibrium. Since the linear
term vanishes and defining $\Delta S=S\left(\vec{\alpha}\right)-S_{0}$
one gets that \begin{equation}
\Delta S=-\frac{1}{2}\sum_{j=1}^{n}\sum_{i=1}^{n}g_{ij}\alpha_{i}\alpha_{j}\label{three}\end{equation}
where the $g_{ij}$'s are equilibrium properties. Using Eqs. (\ref{one})
and (\ref{three}) it follows immediately that the linear laws take
the form:\begin{equation}
\sum_{j=1}^{n}\left(R_{ij}\frac{d\alpha_{j}}{dt}+g_{ij}\alpha_{j}\right)=0\;\;\; i=1,..,n\label{four}\end{equation}
where $R_{ij}=\left(L_{ij}\right)^{-1}$. Assuming that the spontaneous
fluctuations around the equilibrium state of any system obey the same
equations as the macroscopic variables, e.g. Eq. (\ref{four}), using
the methods of statistical mechanics and the fact that the microscopic
equations of motion are invariant under time inversions (microscopic
reversibility), Onsager was able to show that the matrix $L_{ij}$,
the transport coefficients matrix, is symmetric\begin{equation}
L_{ij}=L_{ji}\label{five}\end{equation}
 The proof of this statement may be found in any good textbook in
statistical mechanics \citep{DeGroot3}-\citep{Callen}, or in Refs.
\citep{Onsager1}-\citep{Casimir1}. What is of outmost importance
is to understand that once we admit the validity of Onsager's Reciprocity
Theorem, Eq. (\ref{five}), we are implicitly admitting the linear
regression of the fluctuations hypothesis, which has been thoroughly
corroborated by experiments in colloid systems \citep{Chandra1} \citep{Westgren},
light scattering of fluids \citep{Boon1} and other systems \citep{Berne1}.

However, Casimir \citep{Casimir1} and J. Meixner \citep{Meixner1}
independently pointed out that many irreversible processes did not
fit into Onsager's formulation, in particular the cases where the
state variables are field variables, such as in heat and electrical
conduction. In such cases the state variables satisfy conservation
equations and the forces are given by gradients of intensive variables
instead of defined as in Eq. (\ref{one}). Two major changes appear:
Eq. (\ref{three}) is substituted by the entropy balance equation\begin{equation}
\rho\frac{ds}{dt}+\text{div}\vec{J}_{s}=\sigma\label{six}\end{equation}
which results strictly from the validity of the LEA and the conservation
equations. In Eq. (\ref{six}) $\rho$ is the mass density, $\sigma$
is the entropy production and $\vec{J}_{s}$ is the entropy flux,
which is a measure of the entropy flowing through the boundaries of
the system. Secondly, for an isotropic system $\sigma$ turns out
to be a sum of products of forces and fluxes of the same tensorial
rank. Thus, the linear relationship in Eq. (\ref{two}) is consistent
with $\sigma>0$ but now the fluxes are no longer the time derivatives
of the field variables but the currents. Physically, $\sigma$ is
just the local version of what Clausius in his last paper on thermodynamics
called the \textquotedbl{}uncompensated heat\textquotedbl{}, a measure
of the heat generated by the dissipative effects occurring in a process
taking place in the system \citep{Cropper}. Following the linear
relation hypothesis, the entropy production turns out to be a quadratic
form in terms of the forces, usually written as

\begin{equation}
\sigma=\sum_{i=1}^{n}\sum_{k=1}^{n}L_{ik}X_{i}X_{k}\label{eight}\end{equation}
 such that, if the transport coefficients $L_{ik}$ are positive,
then $\sigma>0$.

Onsager's reciprocity theorem, stated in Eq. (\ref{five}), is kept
as a potulate in the extensions of his theory that constitute nowadays
LIT formalisms. This means that one is ultimately also accepting the
linear regression of fluctuations hypothesis. To summarize this section,
our objection to the use of Eq. (\ref{eq:fr}) comes from the fact
that it is inconsistent with classical irreversible thermodynamics
in two senses: it does not have the structure given by Eq. (\ref{eight})
and it leads to a violation of the linear regression of fluctuations
hypothesis. The proof of this last statement follows directly from
the fluctuaction analysis using standard hydrodynamics techniques.
The full calculation is shown in Ref. \citep{Nos1}.

\section{The Kinetic Theory Approach}

The contradictions that arise from coupling the hydrodymamic acceleration,
or any type of term that is not a gradient of a state variable, with
the heat flux have been clearly stated in the previous section. Since
this coupling seems to arise from following a phenomenological treatment,
one is naturally lead to the question of whether such a relation is
predicted by kinetic theory which, as in the classical case provides
a microscopic framework for the establishment of constitutive equations
from basic principles. The procedure is indeed well-known in the non-relativistic
case through the use of Boltzmann's equation for a dilute gas. In
the relativistic case, as long as the detailed expressions for the
transport coefficients are not required, one can accomplish this task
by using the simplified form of the relativistic Boltzmann equation
known as Marle's equation \cite{ck}. This equation reads, \begin{equation}
v^{\alpha}f_{,\alpha}=-\frac{f-f^{(0)}}{\tau}\label{eq:Marle}\end{equation}
where $v^{\alpha}$ is the molecular four-velocity, $f$ is the single
particle distribution function and $f^{(0)}$ the equilibrium distribution,
i. e. the solution to the homogeneous Boltzmann equation. In Eq. (\ref{eq:Marle})
$\tau^{-1}$ is a parameter which contains the information arising
form the collisions between the particles. However, as mentioned above,
this information is not required for our purposes. Such details are
required in order to calculate transport coefficientes while the structure,
in the sense of the dependence with the state variables, of the dissipative
fluxes can be assesed without them.

The solution to Eq.(\ref{eq:Marle}) can be obtained by the standard
methods of kinetic theory, which have been shown to be valid also
in the relativistic case in Ref. \citep{pa08}. Here we only outline
the procedure since the full calculation has been shown elsewhere
\citep{physica A}. In order to obtain the solution to first order
in the gradients one uses the Chapman-Enskog method in which the distribution
function is proposed as

\begin{equation}
f=f^{\left(0\right)}\left(1+\phi\right)\label{eq:3}\end{equation}
The first term corresponds to the Euler relativistic regime and the
second one, where $\phi$ is the first order correction in the Knudsen
parameter, to the Navier-Stokes level. The equilibrium distribution
function in Eqs. (\ref{eq:Marle}) and (\ref{eq:3}) is a relativistic
Maxwellian which, in the non-degenerate case, is given by \citep{ck,degroor}

\begin{equation}
f^{\left(0\right)}=\frac{n}{4\pi c^{3}z\mathcal{K}_{2}\left(\frac{1}{z}\right)}e^{\frac{u^{\beta}v_{\beta}}{zc^{2}}}\,.\label{eq:4}\end{equation}
where $m$ is the rest mass of the particles, $z=\frac{kT}{mc^{2}}$
the usual relativistic parameter and $\mathcal{K}_{n}$ the modified
Bessel function of the n-th kind. Subtitution of this hypothesis in
Eq. (\ref{eq:Marle}) leads to\begin{equation}
\phi=-\tau v^{\alpha}\left(\frac{\partial f^{\left(0\right)}}{\partial n}n_{,\alpha}+\frac{\partial f^{\left(0\right)}}{\partial T}T_{,\alpha}+\frac{\partial f^{\left(0\right)}}{\partial u^{\beta}}u_{;\alpha}^{\beta}\right)\label{eq5}\end{equation}
The derivatives can be readily calulated and subtituted. The Euler
(previous order) equations are used to write the time derivatives,
appearing in the sum when $\alpha=4$, in terms of the spatial gradients.
Notice that this procedure will naturally lead to a coupling of the
dissipative fluxes, calculated as moments of $\phi$, with the thermodynamic
forces. The solution, in the fluid's comoving frame can be written
as\begin{eqnarray}
f^{(0)}\phi & = & -\tau v^{\ell}f^{\left(0\right)}\left[\frac{n_{,\ell}}{n}+\frac{T_{,\ell}}{T}\left(-1+\frac{\gamma}{z}-\frac{\mathcal{K}_{1}\left(\frac{1}{z}\right)}{2z\mathcal{K}_{2}\left(\frac{1}{z}\right)}-\frac{\mathcal{K}_{3}\left(\frac{1}{z}\right)}{2z\mathcal{K}_{2}\left(\frac{1}{z}\right)}\right)\right]\label{eq:5}\\
 &  & +\tau v^{4}f^{\left(0\right)}\frac{p_{,\mu}h^{\mu\nu}}{c\tilde{\rho}}\frac{v_{\nu}}{zc^{2}}\end{eqnarray}
where $\kappa_{T}$ the isothermal compressibility, $\beta$ the thermal
expansion coefficient and $\tilde{\rho}=\left(n\varepsilon+p\right)/c^{2}$
where $\varepsilon$ and $p$ are the internal energy and pressure
respectively which, in turn, have to still be expressed in terms of
the scalar indepentend variables, $T$ and $n$, through some equation
of state. This is required for consistency, since the chosen representation
is $n$, $T$ and $u^{\nu}$. Then, the forces present in the solution
$\phi$ are exclusively the temperature and density gradients. By
means of the Enskog transport equation \citep{physica A}, the heat
flux in the comoving frame can be shown to be \[
J_{[Q]}^{\mu}=n\left\langle mc^{2}v^{\mu}\right\rangle =mc^{2}\int v^{\mu}f^{(0)}\phi\gamma dv^{*}\]
such that, since $f^{(0)}\phi$ is coupled with $\nabla T$ and $\nabla n$,
the corresponding constitutive equation has the following form \begin{equation}
J_{[Q]}^{\mu}=-h_{\nu}^{\mu}\left(L_{TT}T^{,\nu}+L_{nT}n^{,\nu}\right)\label{eq:fr4}\end{equation}
where $L_{TT}$ is an {}``effective \emph{relativistic} thermal conductivity''.
The new transport coefficient $L_{nT}$ has no classical counterpart.
We would like to remark that equations similar in structure to Eq.
(\ref{eq:fr4}) were proposed by previous authors \citep{ck,degroor,Israel0}
but also to strongly emphasize that the term proportional to the gradient
of $n$ arises here since the pressure $p$ is not an independent
variable, and thus the rules of linear irreversible thermodynamics
compels us to express $\nabla p$ in terms of $\nabla n$ and $\nabla T$.

The fact that kinetic theory predicts a heat flux that depends only
on gradients of state variables is widely accepted. However, as mentioned
above, the phenomenology seems to relate it with the acceleration.
This ambiguety remains an open issue that will be addressed elsewhere.
In the rest of this work we focus on the kinetic theory results given
by Eq. (\ref{eq:fr4}) or variations of it \citep{Israel0,ck,degroor}.

\section{Relativistic Irreversible Thermodynamics: Causality and Stability}

As a natural extension of the arguments summarized in Sect. II, it
is intuitively clear that the formulation of relativistic irreversible
thermodynamics should start by considering systems whose states are
described through field variables which are continuous functions of
the space-time coordinates $x_{\alpha}=\left(x_{1},x_{2},x_{3},ct\right)$
and will hence satisfy conservation equations. If $N^{\alpha}=nu^{\alpha}$
is the particle flux,\begin{equation}
N_{;\alpha}^{\alpha}=0\label{nine}\end{equation}
is the statement of conservation of particles, i. e. the continuity
equation. For the other two state variables, $u^{\nu}$ and $\varepsilon$
(or $T$) the balance equations are no longer independent. Moreover,
the resulting equations must imply the local relativistic versions
of the first and second laws of thermodynamics. The energy-momentum
conservation law\begin{equation}
T_{;\nu}^{\mu\nu}=0\label{twelve}\end{equation}
can be obtained directly from Boltzmann's equation. However, the still
open question is whether this is to be matched with the conservation
law implyied by Einstein's field equation\begin{equation}
R_{\mu\nu}-\frac{1}{2}Rg_{\mu\nu}=\kappa\mathcal{T}_{\mu\nu}\label{eleven}\end{equation}
where $R_{\mu\nu}-\frac{1}{2}Rg_{\mu\nu}$ is Einstein's tensor describing
the geometry of space-time. In Eq. (\ref{eleven}), $\mathcal{T}_{\mu\nu}$
is the energy-momentum tensor which accounts for the properties of
matter and $\kappa$ is the coupling constant. Since the covariant
derivative of the Einstein tensor vanishes one has a conservation
law $\mathcal{T}_{;\nu}^{\mu\nu}=0$. As already shown by Einstein
himself \citep{Einstein1}, the Euler equations of relativistic hydrodynamics
for an inviscid fluid follow directly when\begin{equation}
\mathcal{T}^{\mu\nu}=\frac{n\varepsilon}{c^{2}}u^{\mu}u^{\nu}+ph^{\mu\nu}\label{thirteen}\end{equation}
However, he never address the dissipative case.

On the other hand, from a purely hydrodynamic point of view and in
order to include heat in the total energy account, Eckart resorted
to an irreducible decomposition for $T^{\mu\nu}$ and identified relativistic
corrective terms proportional to the heat flux. His proposal, in his
own words, was that no assumptions would be made for the form of $T_{\mu\nu}$,
but it would be used to define other quantities such as the internal
energy $\varepsilon$ and the heat flux $J_{[Q]}^{\mu}$. He thus
obtained \begin{equation}
T^{\mu\nu}=\frac{n\varepsilon}{c^{2}}u^{\mu}u^{\nu}+ph^{\mu\nu}+\Pi^{\mu\nu\text{ }}+\frac{1}{c^{2}}(J_{[Q]}^{\mu}u^{\nu}+J_{[Q]}^{\nu}u^{\mu})\label{fourteen}\end{equation}
although he never mentioned whether Eq. (\ref{fourteen}) should be
the one compatible with Eq. (\ref{eleven}). One should be cautious
in undertaking this step since heat is not a state variable, but energy
in transit. An alternative to this proposal was examined in Ref. \citep{jnet31}
following Meixner's ideas by omitting the last two terms in Eq. (\ref{fourteen})
and introducing heat in the definition of a total energy flux in a
similar fashion as done in non-relativistic LIT. This approach led
to a debate regarding the question of whether the heat should be included
in Einstein's field equation through the stress-energy tensor \citep{muschik}\citep{reply},
.

In this work as in Ref. \citep{physica A} and Section IV of Ref.
\citep{PRE}, the tensor proposed by Eckart is used while Meixner's
formalism is kept as a separate alternative. Indeed, introducing the
constitutive equation (\ref{eq:fr4}) in the set obtained from (\ref{nine}),
(\ref{twelve}) and (\ref{fourteen}) a closed set of equations is
obtained which will not be repeated here. The next step in order to
examine if Onsager's linear regression of fluctuations hypothesis
holds true implies linearizing such system by setting $T=T_{0}+\delta T$,
$n=n_{0}+\delta n$ and $u^{k}=\delta u^{k}$ ($k=1,\,2,\,3$) since
in equilibrium and in the comoving frame $u_{0}^{k}=0$. Here the
naught subscript denotes equilibrium values and $\delta$ prefix the
fluctuations around them. The linearized set of equations, which the
fluctuations should follow according to the linear regression assumption,
is given by\begin{equation}
\delta\dot{n}+n_{0}\delta\theta=0\label{twentytwo}\end{equation}
 \begin{equation}
\frac{1}{c^{2}}\left(n_{0}\varepsilon_{0}+p_{0}\right)\delta\dot{u}_{\nu}+\frac{1}{\kappa_{T}}\delta n_{,\nu}+\frac{1}{\beta\kappa_{T}}\delta T_{,\nu}-2\eta\left(\delta\tau_{\nu}^{\mu}\right)_{;\mu}-\zeta\delta\theta_{,\nu}-\frac{1}{c^{2}}\left(L_{TT}\delta\dot{T}_{,\nu}+\dot{F}_{\nu}\right)=0\label{twentythree}\end{equation}
 \begin{equation}
nC_{n}\delta\dot{T}+\frac{\beta T_{0}}{\kappa_{T}}\delta\theta-\left[h_{\nu}^{\mu}\left(L_{TT}T^{,\nu}+F^{\nu}\right)\right]_{;k}=0\label{twentyfour}\end{equation}
To arrive at this set of equations the local equilibrium assuption
has been used to express $\varepsilon=\varepsilon\left(n,T\right)$
and $p=p\left(n,T\right)$. $C_{n}$ is the specific heat at constant
particle number density. We conveniently introduced a vector quantity
$F^{\nu}$ as a wildcard which is associated with the hydrodynamic
acceleration in Eckart's fomalism, with a gradient of number density
in the one proposed in Ref. \citep{physica A}, or some other gradient
following Israel's early work \citep{Israel0} or the calculations
shown in relativistic kinetic theory standard books \citep{ck,degroor}.
As was emphatically pointed out in previous work \cite{Nos1}, the
exponential growth of perturbations can be picked up simply by performing
a Fourier-Laplace transform followed by the calculation of the curl
of Eq. (\ref{twentythree}). This isolates the transverse mode and
allows for the assesement of its behavior in time. If $F^{\nu}$ is
associated with a time derivative in the velocity, the procedure yields
a cuadratic equation for the Laplace variable with one postive root
which in turn leads to the unphysical growth of fluctuations. However,
if $F^{\nu}$ is indeed a gradient of a thermodynamic quantity, as
predicted by kinetic theory, its curl vanishes and a first order equation
is obtanied. This yields an exponential decay for the transverse fluctuations
with which one concludes that the instability of the equilibrium state
is not longer present. For the longitudinal mode one obtains

\begin{eqnarray}
\tilde{\rho}_{0}\delta\dot{\theta}+\frac{1}{n\kappa_{T}}\nabla^{2}\delta n+\frac{\beta}{\kappa_{T}}\nabla^{2}\delta T\nonumber \\
-A\nabla^{2}\delta\theta-\frac{L_{TT}}{c^{2}}\nabla^{2}\delta\dot{T}-\frac{1}{c^{2}}\dot{F}_{;\nu}^{\nu} & =0\label{twentyeight}\end{eqnarray}
Equations (\ref{twentytwo}), (\ref{twentyfour}) and (\ref{twentyeight})
constitute a set of three coupled equations for $\delta T$, $\delta n$
and $\delta\theta$. Taking their Fourier-Laplace transform a set
of algebraic equations is obtained which may be solved to yield a
dispersion relation from which its roots may be obtained. After a
rather long calculation by inverse transformation one obtains explicit
solutions for $\delta n\left(\vec{r},t\right)$ and $\delta T\left(\vec{r},t\right)$.

As it is well known the normalized autocorrelation function of the
density fluctuations is proportional to the dynamic structure factor
\citep{Boon1,Berne1,mountain}. This calculation, in the non-relativistic
case, yields the well known Rayleigh-Brillouin spectrum predicted
by Landau and Placzek in 1934 \citep{landau2P} and measured experimentally
for Argon in 1966 by Boon et al. \citep{Boon1}. This is one of the
many experimental confirmations of the linear regression assumption.

What we expect to find in the relativistic case is precisely a relativistic
correction to this spectrum. This does not occur when if $F^{\mu}\propto\dot{u}^{\mu}$.
As shown in Ref. \cite{PRE}, the spectrum simply does nor exist.
On the other hand, if $F^{\mu}\propto h^{\mu\nu}n_{,\nu}$ we do indeed
recover the spectrum with relativistic corrections to Rayleigh's peak,
which in the non-relativistic limit reduces to the classical form.
Even if the experiment is not or cannot be performed for technological
reasons, the result speaks in favor of a constitutive equation in
the form given in Refs. \cite{physica A,degroor,ck,Israel0} and not
Eckart's one. It is precisely due to the structure of Eq. (\ref{eq:fr4})
that one can show that the resulting linearized relativistic hydrodynamic
equations are in full agreement with Onsager's linear regression assumption
\citep{Onsager1} \citep{DeGroot2}. This in turn implies that the
equilibrium state is thermodynamically stable.

\section{Discussion}

In the previous sections, we have reviewed recent results in relativistic
LIT and kinetic theory and pointed out some key theoretical arguments
related to them. In Sect. II, we emphasized the meaning of the generic
instabilities found in Ref. \citep{HL} as violations of Onsager's
linear regression assumption. This hypothesis, as dicussed above,
lies deep inside the theoretical setup of LIT and its violation is
reason enough to discard a theory of irreversible processes. Based
on this violation by Eckart's linear theory, higher order formalisms
were developed and now generally favored even though they also have
been objected \citep{LGCS1,LGCS2}. However, based in our recent results
and as outlined here in Sects. III and IV, the source of this unphysical
behavior is the coupling of heat with acceleration assumed by Eckart's
theory which is in disagreement with standard kinetic theory. These
facts do not rule out the possibility of considering higher order,
or extended, theories \citep{israel1,israel2,jou} but questions the
need for them. That is, the unphysical behavior found in Eckart's
formalism is corrected once kinetic theory is used in order to construct
a heat flux tensor. These findings reopen the possibility of modeling
relativistic fluids with first order theories.

In Sect. II we outlined the calculation of Ref. \citep{physica A}
where we obtained a constitutive equation that couples the heat flux
with the thermodynamic forces $\nabla T$ and $\nabla n$. This is
clearly consistent with the generalization of Onsager's theory since
it couples fluxes with thermodynamic forces of the same tensorial
rank, the former in this case being the heat flux and the latter $h^{\mu\nu}T_{,\nu}$
and $h^{\mu\nu}n_{,\nu}$ in the $(n,T,u^{\mu})$ representation.
This constitutive equation is then introduced in Sect. IV to argue
that with it no stability nor causality issues arise in the relativistic
transport equations. The details of these calculations can be found
in Refs. \citep{Nos1} and \citep{PRE}. We also comment on the causality
and refer the reader to a recent publication in this subject \cite{ante}.

This work also serves to update a discussion that has been going on
for some time. The first issue concerns the nature of the time component
of heat flux four-vector. It is important to notice that we are using
a projector in the constitutive equation (\ref{eq:fr4}). This was
questioned in previous work \citep{jnet31} but has since been analyzed.
Indeed, in Ref. \citep{pa08} we show that a generalized relativistic
Chapman-Enskog procedure for the relativistic Boltzmann equation is
possible and consistent with irreversible thermodynamics. We concluded
that the fourth component of the heat flux as predicted by such a
formal approach is still the internal energy itself. Thus, no {}``disspation
in the time direction'' is present in the theory and the projector
should be included in the constitutive equation as done by most authors.

Secondly, the presence of heat in the stress-energy tensor must be
addressed here following the discussion in Refs. \citep{muschik}
and \citep{reply}. In Sect. IV, in particular in Eq. (\ref{fourteen}),
we clearly state that here as in some of our recent work, the stress-energy
tensor considered is indeed the one proposed by Eckart strictly from
a mathematical argument. However it still lacks a kinetic justification
in the laboratory frame, eventhough some work along this line can
be found in Ref. \citep{AAAA}. Nevertheless the relativistic heat
terms can be readily shown to be present in the tensor in the comoving
frame where most of our calculations are performed. Indeed we have
lately adopted such a tensor as the fluid's stress energy tensor where,
as mentioned earlier, the heat flux is included and the first term
corresponds to an internal energy flux. Both energy and momentum balances
are obtained from the general conservation law $T_{;\nu}^{\mu\nu}=0$
as pointed out above. On the other hand, the work in Ref. \citep{jnet31},
which raised the discussion leading to Refs. \citep{muschik} and
\citep{reply}, used a Meixner-like approach where the heat is not
included in the stress energy tensor but is introduced in the system
via the construction of a total energy flux. The predictions found
in \citep{PRE} finally lead to the criteria which, on experimental
grounds, should prevail to decide which of the approaches is the correct
one: Eckart's \citep{Eckart1}, the one proposed by Sandoval-Villalbazo
and Garcia-Colin in Ref. \citep{jnet31}, or the one in Ref. \citep{physica A}.
This is still an open question.

It has to be clarifyed here that in the present work as well as in
Refs. \citep{physica A} and Section IV of Ref. \citep{PRE}, the
tensor given in Eq. (\ref{fourteen}) is used as the fluid's stress
energy tensor eventhough a kinetic proof of its structure, other than
in the commoving frame, is still to be established. We consider that
the question of whether this same tensor is the one to be included
in Einstein's field equation is still open and should be addressed.
The actual solution of such equation including heat sources should
be critically analyzed in order to asses the possible effects of including
such terms. This is clearly outside the scope of this review and will
be addressed elsewhere.

The authors deeply acknowledge the valuable comments and suggestions
of Prof. L.S. Garcia-Colin.

\end{document}